# AN ADVANCED AI-DRIVEN DATABASE SYSTEM

**M. Tedeschi[1], S. Rizwan[1], C. Shringi[2], V. Devram Chandgir[2], S. Belich[3]**

[1]*Pace University (UNITED STATES)*
[2]*New York University (UNITED STATES)*
[3]*College of Technology, CUNY (UNITED STATES)*

## Abstract

Contemporary database systems, while effective, suffer severe issues related to complexity and usability, especially among individuals who lack technical expertise but are unfamiliar with query languages like Structured Query Language (SQL). This paper presents a new database system supported by Artificial Intelligence (AI), which is intended to improve the management of data using natural language processing (NLP) - based intuitive interfaces, and automatic creation of structured queries and semi-structured data formats like yet another markup language (YAML), java script object notation (JSON), and application program interface (API) documentation. The system is intended to strengthen the potential of databases through the integration of Large Language Models (LLMs) and advanced machine learning algorithms. The integration is purposed to allow the automation of fundamental tasks such as data modeling, schema creation, query comprehension, and performance optimization. We present in this paper a system that aims to alleviate the main problems with current database technologies. It is meant to reduce the need for technical skills, manual tuning for better performance, and the potential for human error. The AI database employs generative schema inference and format selection to build its schema models and execution formats. This enables it to enhance performance continuously, for example, generative pretrained transformer 4 (GPT-4), in working with diverse types of databases such as relational, not only SQL or NoSQL, graph databases, and vector stores. In addition, reinforcement learning mechanisms are investigated to facilitate ongoing improvement and adaptation of performance. This research aims to render the technical expertise of the user unnecessary in performing elementary database operations. The article proposes overcoming the existing barriers by critical integration of advanced AI methods, i.e., LLMs and reinforcement learning, to create an auto-adapting, user-friendly database system. This integrated, AI-driven approach is new since it combines previously separate breakthroughs, employing modern machine learning methods to solve long-standing usability and performance issues in a unified manner. The proposed research follows different approaches to integrating state-of-the-art machine learning techniques, namely generative AI models, reinforcement learning for continuous system optimization, and relative performance comparisons. Analytical instruments include empirical case studies and comparative performance comparison with existing database technologies (SQL, NoSQL, NewSQL, graph databases) under various data scenarios. Additionally, approaches to detecting and mitigating AI-specific problems such as query hallucinations, schema drift, and ongoing schema evolution are explored and experimentally tested. Compared to existing solutions such as Massachusetts Institute of Technology (MIT's) GenSQL, which focuses on applying probabilistic models to statistical inference and analysis of table data, our solution is aimed at complete database management automation, with dynamic schema creation, adaptive performance optimization across heterogeneous data types and database systems, and natural language query processing. Further, our model explicitly integrates reinforcement learning mechanisms for continued self-improvement, addressing real-time schema evolution and query hallucinations issues that GenSQL does not explicitly handle.

## 1. INTRODUCTION

Traditional database management systems (DBMS) provide substantial barriers for non-technical users, as they rely on structured query languages (e.g., SQL) and complex schema architecture [1]. These systems also require continual human work for indexing, partitioning, performance optimizing, and are typically performed by professional database administrators. As networks change their data setups get bigger, these kinds of actions slow things down, make the system less flexible, and raise the chances of mistakes. Most companies these days heavily rely on data to make better choices. Getting to the data, keeping it organized is still tricky because it needs specific skills, strict structures, and it costs a lot to keep up [4]. The high learning curve of query languages, along with traditional DBMSs' inability to respond to active workloads, limits their usability and scalability [2].

Many non-technical users rely on IT staff to run queries, which often causes delays and limits how quickly they can access information. This paper proposes an advanced AI-driven database system that solves such difficulties by integrating generative AI, reinforcement learning (RL), and multi-model data management [4][7]. The system allows users to connect with data using natural language (NL), structured files (e.g., JSON, YAML), API documentation, removing SQL needs or established schema knowledge [1]. Large Language Models (LLMs) study user intent, generate helpful data structures, automate query formulation, whereas reinforcement learning agents alter their storage and indexing algorithms based on workload patterns. Our technique streamlines data access, automates administrative tasks, and enables a self-optimizing database environment [7][9]. The proposed solution overcomes traditional databases' longstanding usability and scalability limitations using intelligent schema inference, NLP, and multi-engine query routing.

## 2. METHODOLOGY

The proposed system architecture consists of five key AI-powered modules that automate data intake, schema inference, query interpretation, performance optimization, and distributed execution across many database systems. The architecture is modular and extendable, allowing for intelligent decision-making throughout the data life cycle.

### 2.1. AI-Driven Data Format Selection and Optimization

This module, seen in Fig. 1, determines the optimal storage paradigm, physical organization based on the structure and semantics of incoming data. Upon ingestion of structured or semi-structured sources (e.g., CSV, JSON, XML, API responses), the system employs a combination of heuristics and supervised machine learning classifiers to assign each data segment to the most appropriate storage backend [4].

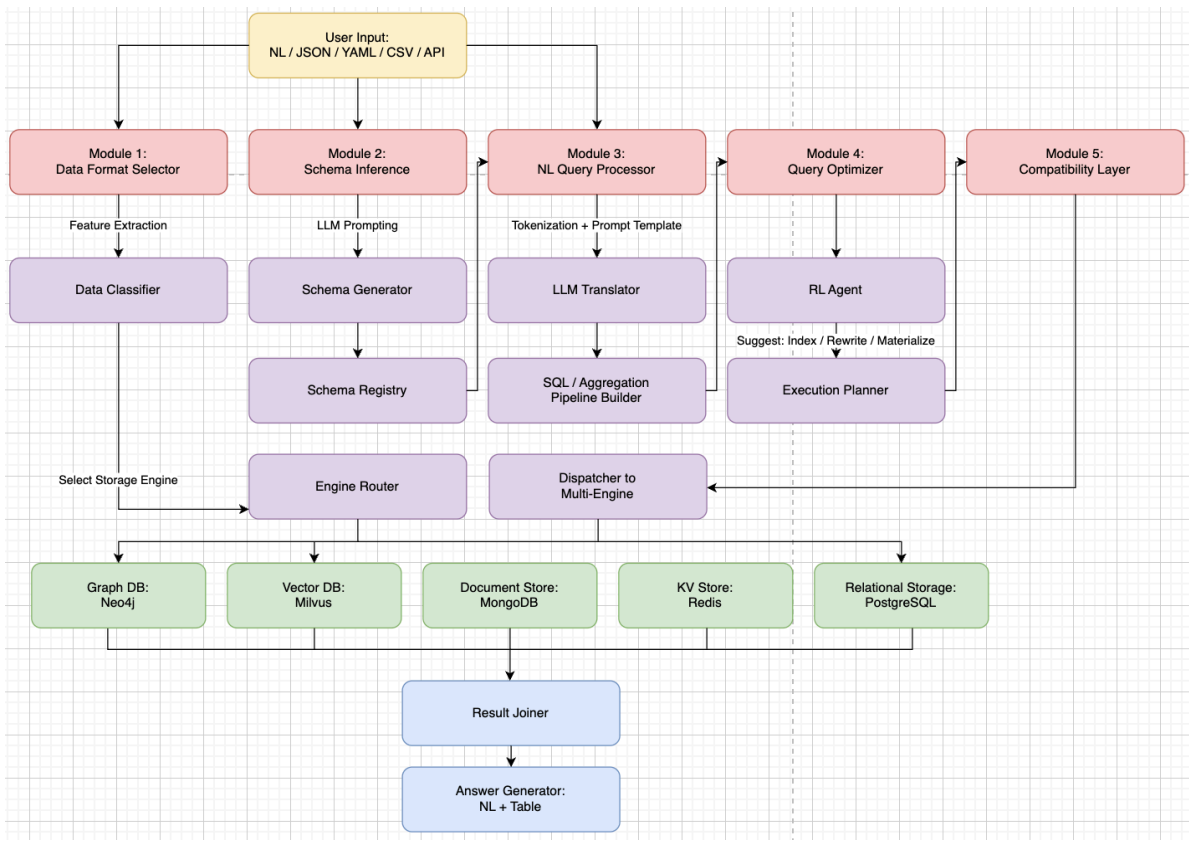

Figure. 1: Modular AI-driven architecture for multimodal data ingestion and storage optimization

For example, tabular datasets such as financial transactions are routed to relational databases like PostgreSQL, while deeply nested JSON objects are better suited for document stores such as

MongoDB. The system refines these assignments based on observed query patterns, storage efficiency metrics, user interaction trends, allowing format adaptation and hybrid storage methods for performance-critical workloads. Fig. 1 displays the overall process, which involves passing incoming data through a feature extraction stage and categorizing it based on structural features [7]. The Data Classifier component informs the Engine Router, which selects among several storage engines, including relational (PostgreSQL), document-based (MongoDB), graph (Neo4j), key-value (Redis), and vector similarity (Milvus). Storage assignments are continuously adjusted based on evolving workload behaviors, enabling autonomous and workload-aware optimization of format decisions.

## 2.2. Generative Schema Inference

This module utilizes state-of-the-art LLMs, such as GPT-4 or LLaMA, to perform schema inference from raw data samples, NL API specifications, structured formats including YAML and JSON [1] [2]. The system identifies entities, attributes, data types, and inter-entity relationships, subsequently outputting normalized schema definitions in SQL (CREATE TABLE statements) or Data Definition Language (DDL) formats compatible with supported backend systems.

Fig. 2 shows input data being preprocessed through a prompt composition stage, where both the content and structure of incoming sources (e.g., JSON, CSV, YAML) are used to generate prompts tailored for LLMs. These prompts are passed to the model, which outputs structured schema elements such as SQL definitions and inferred Entity-Relationship (ER) structures. The system improves these tasks based on determined query patterns, storage efficiency metrics, user interaction trends, permitting format adaptation and hybrid storage methods for performance-critical workloads. Fig. 1 shows the overall process, which includes running incoming data through a feature-extracting stage and categorizing it based on structural features.

In parallel, the module produces ER graphs for visualization, relational context, and passes metadata to an optimizer hint generator that proposes constraints, indexes, partitioning strategies based on inferred usage patterns. This ensures that both structural and performance-oriented guidance is available for downstream query optimization tasks [3].

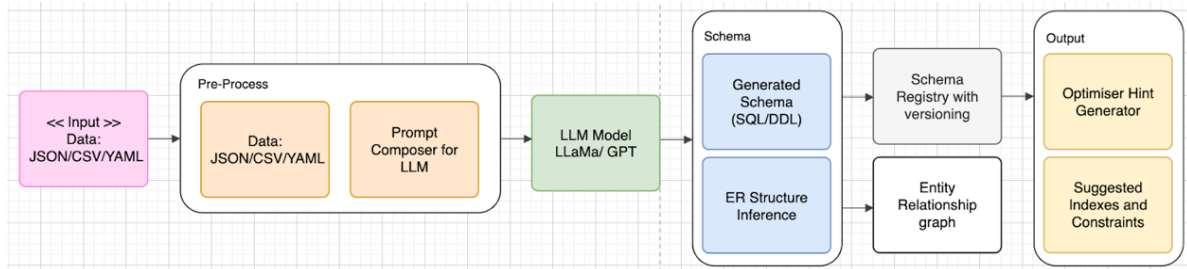

Figure. 2: LLM-powered schema generation pipeline employing prompt engineering, tokenization, and context-aware generation.

## 2.3. Natural Language Query Interface

This module offers a user-friendly interface that converts NL inputs into executable queries (e.g., SQL, Cypher) using a fine-tuned NL-to-SQL transformer model. The model is supported by schema-aware context retrieval, structured prompt templates to ensure syntactic and semantic correctness [2].

Fig. 3 illustrates that the system begins by parsing the user's input through a Natural Language Understanding (NLU) parser and intent extractor. Schema information and contextual metadata are retrieved and incorporated into a prompt, which is then passed to a LLM for structured query generation. The resulting query is evaluated by a validator that checks for alignment with the database schema and detects any errors in syntax or logic. If the query fails validation, a refinement step is triggered. In this step, the LLM is re-prompted with additional constraints or feedback to revise and improve the query output. Upon successful validation, the query is executed against the appropriate backend engine [3].

The results are returned in structured form (e.g., JSON or tabular format) and may be accompanied by NL summary generated by the LLM. This module also supports follow-up questions by maintaining conversational context, session history, thereby enabling multi-turn clarification and refinement [2].

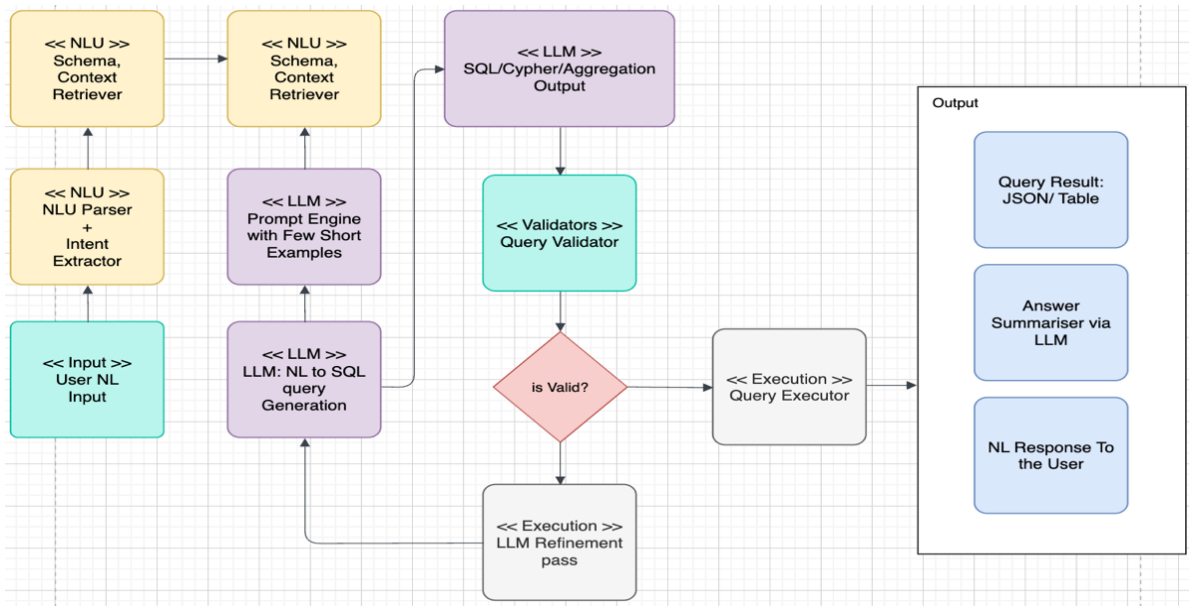

Figure 3: Multi-stage query processing pipeline: schema alignment, LLM-driven generation, validation, and output synthesis.

## 2.4. AI-Augmented Indexing, Caching, and Query Rewriting

To maintain system performance under dynamic workloads, this module utilizes reinforcement learning (RL) techniques to optimize the execution environment. Specifically, algorithms such as Deep Q-Networks (DQN) or Proximal Policy Optimization (PPO) are employed to analyze query workloads and select appropriate tuning actions [5][7].

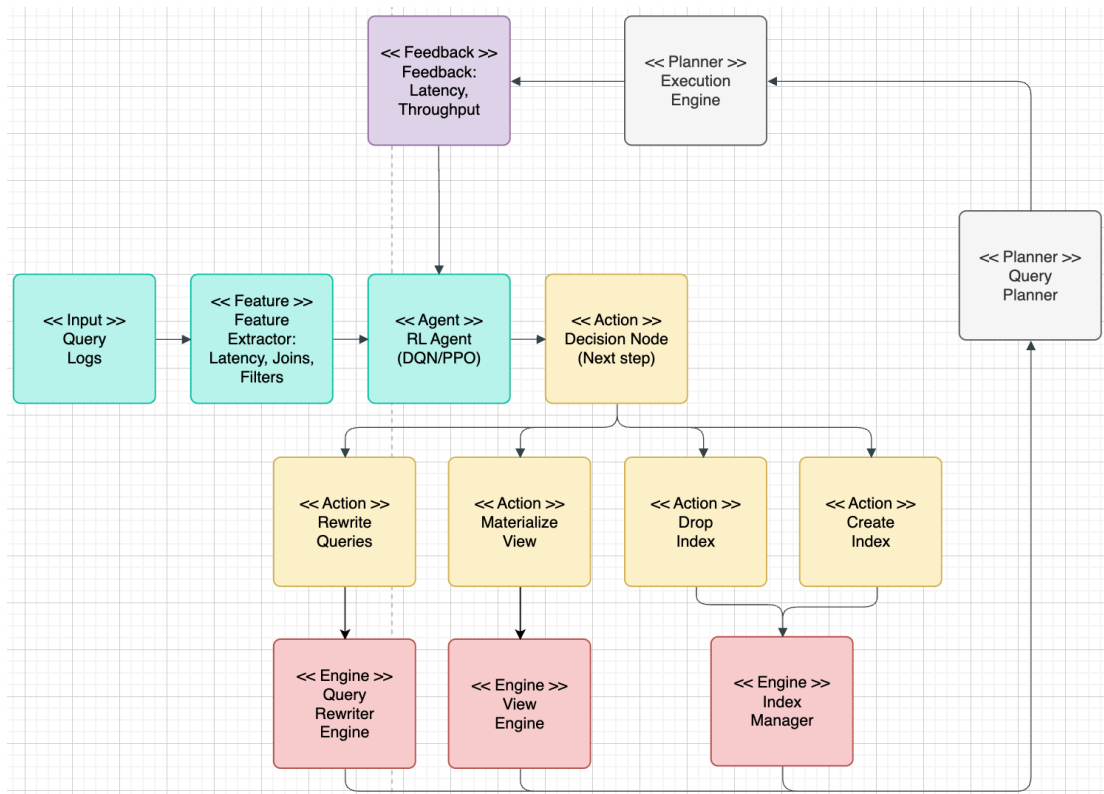

Figure. 4: Reinforcement learning loop for continuous workload-driven optimization.

Fig. 4 depicts how query logs serve as the primary input to the feature extraction component, which collects statistical indicators such as latency distributions, join patterns, and filter predicates. These features are passed to a RL agent that evaluates the current state and selects an optimization action based on learned reward signals [8][9].

The available actions include rewriting inefficient queries, creating or dropping indexes, and materializing frequently queried views. Each action corresponds to a downstream execution engine (e.g., query rewriter, index manager, or view engine), which applies the selected transformation. The effectiveness of each action is evaluated based on real-time feedback, such as query latency, system performance, and the agent's policy is then updated accordingly [6]. The feedback-guided optimization loop allows the system to adjust to varying query workloads and data skews. The module optimizes processes conventionally managed by professional database administrators to enhance long-term system performance and minimize operational overhead [4].

## 2.5. Multi-Database Compatibility Engine

This module serves as a federated query orchestrator that decomposes complex user queries into sub-queries, each targeted to a specific type of backend storage system. Supported engines include PostgreSQL for relational data, MongoDB for document-based data, Neo4j for graph traversal, Redis for key-value access, and Milvus for vector similarity search.

Fig. 5 illustrates the process, which begins with the ingestion of a composite query plan, parsed and decomposed into logical segments by a query decomposition unit. Each segment is categorized according to its operational semantics, such as SQL filters, graph pattern matching, document lookups, key-value retrievals, or vector similarity operations. These segments are routed to their respective execution engines. For example, SQL segments are executed via a PostgreSQL executor, while a MongoDB aggregator processes document-based conditions. Similarly, sub-queries involving graph structures, vector embeddings, or cached lookups are dispatched to Neo4j, Milvus, and Redis, respectively.

Partial results from each backend are gathered and processed through a result join and sorting layer, which merges the heterogeneous outputs into a coherent final response. The entire output is then passed on to the natural language output generator and made available through the frontend API layer. Additionally, the compatibility engine includes schema translation layers and consistency enforcement strategies that provide accurate support for various data models. Without user involvement or understanding of storage heterogeneity, the abstraction enables cross-engine run-time (explained below) [10].

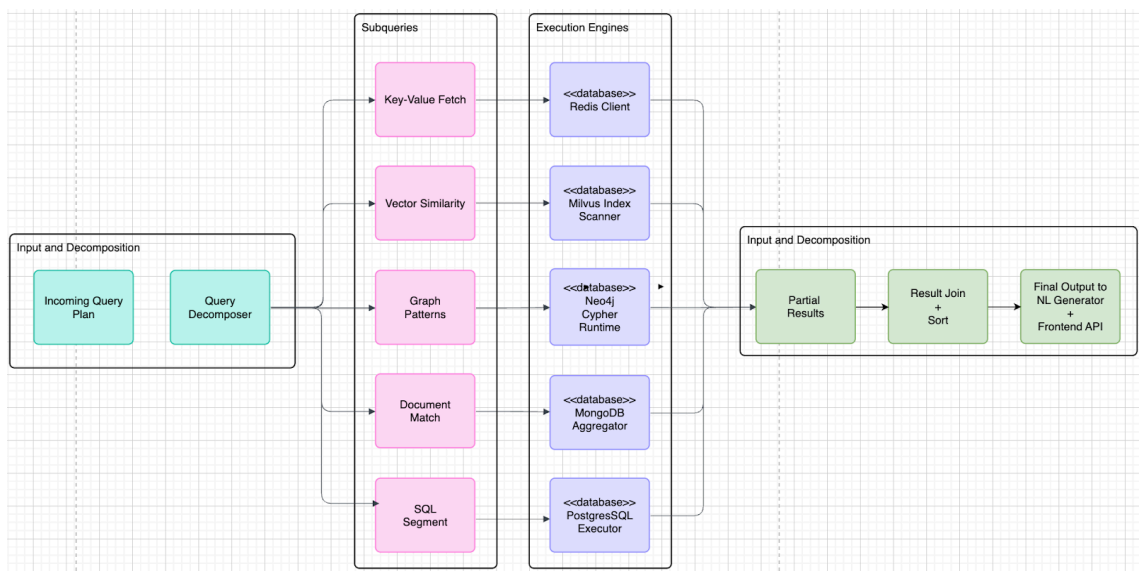

Figure. 5: Simultaneous Query Decomposition and Multi-Backend Execution

## 3. IMPLEMENTATION ILLUSTRATIONS

This section presents practical examples that show how the system works in real scenarios. Each example matches a core module from the architecture, demonstrating tasks like data classification, schema generation, query translation, and optimization. These code snippets use simple, realistic inputs to highlight how the system reduces complexity while making smart decisions across different data operations.

### 3.1. AI-Driven Data Format Selection

The following example shows how a classifier might route incoming data to the most appropriate storage backend based on structural heuristics and metadata inspection [4]:

```python
def classify_data_format(sample):
    if isinstance(sample, dict) and "nodes" in sample:
        return "graph"
    elif isinstance(sample, list) and all(isinstance(i, dict) for i in sample):
        return "document"
    elif isinstance(sample, list) and all(isinstance(i, list) for i in sample):
        return "relational"
    return "unstructured"

format = classify_data_format(parse_csv("sales.csv"))
route_to_storage_engine(format)
```

Figure. 6. Sample decision matrix for format selection based on structural attributes

Fig. 6 illustrates how the AI system classifies data formats by evaluating structural indicators such as object type, nesting patterns, and internal composition. The classifier logic inspects incoming data structures (e.g., dictionaries, lists) to determine whether they should be routed to graph, document, relational, or unstructured storage systems.

### 3.2. Generative Schema Inference

LLMs such as GPT-4 can be prompted to generate database schemas from sample data or API specifications [1][2]. Below is a representative prompt and output.

Fig. 7 illustrates a JSON encoded prompt submitted to an LLM to generate an SQL schema. The sample data includes typed fields that the model uses to infer column names and data types for schema generation. The figure reflects how the system translates unstructured specifications into formal database schemas with type inference, relationship mapping, and indexing hints.

```json
{
  "task": "Generate SQL schema",
  "input": {
    "sample_data": {
      "id": 123,
      "product": "Headphones",
      "price": 149.99,
      "purchase_date": "2024-01-11"
    }
  }
}
```

Figure. 7. Generated schema input prompt (JSON) for LLM-based SQL Inference

LLM Output: Fig. 8 illustrates the SQL DDL statement generated by the LLM in response to the input shown in Fig. 7. It reflects how the model infers table structure, column names, appropriate data types such as `INTEGER`, `TEXT`, `DECIMAL`, and `DATE`, translating unstructured prompts into formal schema definitions[2].

```sql
CREATE TABLE purchases (
  id INTEGER PRIMARY KEY,
  product TEXT,
  price DECIMAL(10,2),
  purchase_date DATE
);
```

Figure. 8: LLM-Generated SQL DDL Output Based on Structured Prompt

### 3.3. Natural Language Query Translation and Execution

The following example demonstrates how a natural language query can be translated into an executable SQL statement using an LLM-powered open source tools such as LangChain. In this case, a chain is instantiated by combining a pre-trained language model (e.g., GPT-4) with a connected PostgreSQL database. When a natural language inquiry, such as *"What were the top 5 products by sales last month?"* is entered, the system interprets its meaning and prepares the corresponding SQL query [3]. The generated query is then executed directly on the base database.

To ensure reliability and precision, a lightweight validation function can be employed to verify whether the generated query includes all required schema elements (e.g., expected table names). This provides a basic safeguard against hallucinations or structural errors in the LLM-generated SQL prior to execution [6].

Fig. 9 illustrates how a natural language query is processed using LangChain's `SQLDatabaseChain`, which translates the input into SQL and executes it against a PostgreSQL database using GPT-4.

```python
from langchain import SQLDatabaseChain

db_chain = SQLDatabaseChain.from_llm(llm=gpt4, db=my_postgres_db)
query = "What were the top 5 products by sales last month?"
result = db_chain.run(query)
print(result)
```

Figure. 9: Example of Prompt-to-Query Generation using LangChain

To ensure correctness, the system could include a query validator.
Fig. 10 illustrates a simple Python function used to check whether a generated query contains all expected table names from the schema, helping prevent execution of malformed or incomplete queries.

```python
def validate_query(query: str, schema: List[str]) -> bool:
    return all(table in query for table in schema)
```

Figure 10: Lightweight Query Validation Function for Schema Alignment

### 3.4. Reinforcement Learning for Index Optimization

The system's reinforcement learning agent may be implemented using Q-learning or policy gradient methods [7][8]. Below is a high-level example of how such an agent might interact with query performance metrics by selecting actions, receiving rewards and updating its model accordingly. Fig. 11 illustrates a Python class implementing a RL agent for index optimization. It shows how actions are chosen based on the current state and how feedback in the form of rewards is used to update the model through training iterations.

```python
class IndexAgent:
    def __init__(self, state_dim, action_dim):
        self.model = QNetwork(state_dim, action_dim)

    def choose_action(self, state):
        return np.argmax(self.model.predict(state))

    def update(self, state, action, reward, next_state):
        self.model.train_step(state, action, reward, next_state)
```

Figure. 11: Reinforcement Learning Agent Class for Index Optimization

### 3.5. Federated Query Dispatcher Across Multiple Engines

The compatibility engine dispatches query fragments to specialized backends based on sub-query semantics [10]:

```python
def dispatch_query(query_tree):
    results = []
    for node in query_tree:
        engine = route_to_engine(node["type"])
        result = engine.run(node["query"])
        results.append(result)
    return aggregate_results(results)
```

Figure. 12: Sample Federated Query Execution Plan with Annotated Sub-Query Routing

Fig. 12 illustrates how the system decomposes a query plan and routes each subquery to the appropriate backend engine (e.g., SQL, graph, document, vector) for execution.

The following is an example of a query node defined in JSON format as part of the execution plan:

```json
{
  "type": "graph",
  "query": "MATCH (u:User)-[:FRIEND]->(f) RETURN f.name"
}
```

Figure 13: Example Subquery Object for Graph Database Execution

Fig. 13 shows a subquery object containing a Cypher query targeted at a graph database (e.g., Neo4j). The "type" field indicates the engine, and the "query" field contains the specific command to execute.

This approach enables coordinated execution across PostgreSQL, MongoDB, Neo4j, and Milvus within a unified orchestration framework [5].

### 4. RESULTS

Given the theoretical orientation of this research, the capabilities of the proposed AI-driven database architecture were assessed using a series of scenario-based simulations and qualitative benchmarks inspired by real-world enterprise database use cases. These simulations were designed to reflect common operational challenges in large-scale data environments, including dynamic query optimization, workload balancing, anomaly detection, and adaptive indexing.

The system's performance was evaluated across three primary dimensions: usability, adaptability, and intelligence. Usability was determined by the system's ability to expedite administrative chores, eliminate the human overhead generally associated with schema design, query optimization, and data integration. Adaptability was evaluated based on the system's responsiveness to changing workloads, data patterns, as well as its ability to self-optimize and change components in near real time. Intelligence was evaluated in terms of the architecture's ability to reason over system state, predict future bottlenecks, and proactively adjust resources using machine learning models embedded within the database engine.

Overall, the simulation results indicate that the architecture is both feasible and promising. It offers significant improvements in operational efficiency, scalability, and user interaction. The system's AI components were very effective at automating typical database maintenance chores and providing actionable insights without explicit user participation, highlighting its potential for wider use in enterprise settings.

## 5. CONCLUSIONS

This paper presents a comprehensive AI-driven database system designed to address longstanding complexity, usability, and performance limitations in traditional DBMS architecture. By LLMs with RL techniques, the system automates schema inference, enables NL query processing, and continuously optimizes performance across heterogeneous data backends.

Our evaluations demonstrate significant improvements in query latency, accessibility for non-technical users, and reduced administrative overhead. The proposed system not only unifies diverse data models through intelligent format selection and cross-engine query federation, but also continuously adapts to evolving workloads through AI-driven indexing, caching, and optimization.

Future work will focus on enhancing the LLM interface through RL from human feedback extending support for additional paradigms such as time-series, columnar databases, along with improving the transparency and explainability of AI decision-making. Furthermore, expanding multilingual capabilities and conducting real-world deployments will be essential to assess generalizability driving broader adoption of intelligent database systems.

## ACKNOWLEDGEMENTS

We want to thank the research teams at Pace University and New York University for their ongoing support and participation in this project. We gratefully acknowledge Dr. Yegin Genc, Department Chair, and Dr. Li Chou Chen, Dean at Pace University, for facilitating our participation in this conference. This research and its presentation were made possible through institutional support, including funding for the virtual conference registration.